\documentclass[letterpaper]{article} 
\usepackage[submission]{aaai23}  
\usepackage{times}  
\usepackage{helvet}  
\usepackage{courier}  
\usepackage[hyphens]{url}  
\usepackage{graphicx} 
\usepackage{natbib}  
\usepackage{caption} 
\usepackage{algorithm}
\usepackage{algorithmic}
\usepackage{todonotes}
\usepackage{amsfonts}
\usepackage{hyperref}
\usepackage{kbordermatrix}
\usepackage{booktabs}
\usepackage{tikz}
\usepackage{tikz-cd}
\usepackage{amsmath}
\usepackage{amsfonts}
\usepackage{latexsym}
\usepackage{enumerate}
\usepackage{wasysym}
\usepackage{lipsum}
\usepackage{amssymb}
\usepackage{amsthm}
\usetikzlibrary{positioning,decorations.pathreplacing,angles,quotes}
\newcommand{\N}{\mathcal{N}}
\newcommand{\C}{\mathcal{C}}

\newtheorem{example}{Example}

\usepackage{newfloat}
\usepackage{listings}
\DeclareCaptionStyle{ruled}{labelfont=normalfont,labelsep=colon,strut=off} 
\floatstyle{ruled}
\newfloat{listing}{tb}{lst}{}
\floatname{listing}{Listing}

\usepackage{amsfonts,amsmath,amsthm}
\usepackage{cleveref}
\usepackage{enumerate}
\usepackage{bbm}
\usepackage[inline]{enumitem}

\theoremstyle{definition}

\title{Selecting Representative Bodies: An Axiomatic View \footnote{This project was conducted at the Algorithmic Technology for Democracy workshop, organized by the Lorentz Center (Oort) and Davide Grossi, Ulrike Hahn, Michael Maes and Andreas Nitsche.}}
\author {
    Manon Revel,\textsuperscript{\rm 1, \rm 2}
    Niclas Boehmer, \textsuperscript{\rm 3}
    Rachael Colley, \textsuperscript{\rm 4}\\
    Markus Brill, \textsuperscript{\rm 5}
    Piotr Faliszewski, \textsuperscript{\rm 6}
    Edith Elkind \textsuperscript{\rm 7}
}
\affiliations {
    \textsuperscript{\rm 1} Massacchussetts Institute of Technology,
    \textsuperscript{\rm 2} Harvard University,
    \textsuperscript{\rm 3} TU Berlin,
    \textsuperscript{\rm 4} Université Toulouse Capitole,\\
    \textsuperscript{\rm 5} University of Warwick,
    \textsuperscript{\rm 6} AGH University,
    \textsuperscript{\rm 7} University of Oxford\\
     
    mrevel@mit.edu, niclas.boehmer@tu-berlin.de, rachael.colley@irit.fr, \\ markus.brill@warwick.ac.uk, faliszew@agh.edu.pl, elkind@cs.ox.ac.uk
}

\usepackage{bibentry}

\begin{document}

\maketitle
\begin{abstract}
  As the world's democratic institutions are challenged by dissatisfied citizens, political scientists and also computer scientists have proposed and analyzed various (innovative) methods to select representative bodies, a crucial task in every democracy. However, a unified framework to analyze and compare different selection mechanisms is missing, resulting in very few comparative works. 
To address this gap, we advocate employing concepts and tools from computational social choice in order to devise a model in which different selection mechanisms can be formalized. Such a model would allow for desirable representation axioms to be conceptualized and evaluated.
We make the first step in this direction by proposing a unifying mathematical formulation of different selection mechanisms as well as various social-choice-inspired axioms such as proportionality and monotonicity.
\end{abstract}

\section{Introduction}
It is often argued that representative democracy is in crisis (e.g., see Chapter 2 in the book by  \citet{landemore2020open} and the references therein). In particular, the justification of \textit{representative bodies} is called into question whenever they make decisions that appear to go against the interests of those they are supposed to represent.
In line with this, a survey by the Pew Research Center \citep{pew2} finds that, while there remains broad global support for representative democracy, there is also a strong sense that existing political systems need reform. 

In this paper, we focus on the task of selecting a representative body, which is a crucial ingredient of all democratic institutions as argued by political scientists \cite{mill1924representative,mansbridge2009selection,landa2020representative,rae1967political,lijphart1990political,michener2021remoteness}. 
There is no shortage of innovative proposals to change how representative bodies are selected around the world. 
For example, some propose
to select representatives at random (a.k.a. \textit{sortition}) \citep{bouricius2020democracy},  to elect them through transitive delegations (a.k.a. \textit{liquid democracy}) \citep{valsangiacomo2021political}, or to drastically increase the size of parliaments (see, e.g., \url{https://thirty-thousand.org}). Each proposed method has its benefits and drawbacks; however, we lack a systematic way to evaluate and compare them.
Specifically, while there are numerous works in computer science and political science analyzing the strengths and weaknesses of specific methods, principled comparisons are rare.%
\footnote{The few existing exceptions mostly focus on epistemic aspects, the robustness of representation, and majority agreement \cite{alouf2022should,Gree15a,gelauffrepresentational,abramowitz2019proxy}.}

We call for the development of a unified framework to formulate and compare innovative and traditional selection mechanisms on a more principled basis. 
Having formulated different mechanisms within the same framework then offers the possibility to formulate different desiderata in the same framework.
While we believe that comparisons from different perspectives are possible and, in fact, urgently needed, we put forward an \textit{axiomatic} view on selection mechanisms, drawing inspiration from the rich social choice literature on voting rules.
In the main part of the paper, we give a concrete example of how a systematic comparison from an axiomatic perspective of selection mechanisms could look like.  
Firstly, we present a simple yet rich mathematical framework to formulate different selection mechanisms. 
Secondly, we define various axioms capturing notions of \textit{cogent representation}. 
We hope that these axioms can be used to quantitatively investigate inherent and poorly understood trade-offs at the heart of democratic innovations.
Notably, our research program's focus is not on finding the ``ideal'' representation system. We rather envision building a navigator that maps selection mechanisms to axioms. 
We advocate for building a coherent picture of the advantages and disadvantages of competing selection proposals to gear public debates towards \textit{what kind of trade-offs societies are facing}, instead of continuing to argue for competing selection mechanisms on disconnected grounds.

In recent years, computer science and  democratic innovations have become increasingly intertwined, with computer scientists tackling many algorithmic design and scalability problems arising in different representation schemes and analyzing such schemes axiomatically.
In fact, only with advancements in information technology, the idea of more complex and interactive voting models is becoming more commonplace \cite{brill2018interactive}. 
\citet{miller1969program} and \citet{tullock1992computerizing}, for instance, argued that richer political decision-making processes on a nationwide scale have recently become possible thanks to technology. 
Our envisioned research program again relies on the expertise of computer scientists. 
More specifically, many areas of the AAMAS community could contribute to our endeavour to mathematically formulate and analyze selection mechanisms and desiderata. 
For instance,
\begin{enumerate*}[label=(\roman*)]
\item the \textit{Social Choice and Cooperative Game Theory} community has expertise in the axiomatic analysis of voting rules,
\item the \textit{Coordination, Organisations, Institutions, and Norms} community can contribute a normative perspective, and 
\item the \textit{Humans and AI / Human-Agent Interaction} area could help in the analysis of usability aspects.
\end{enumerate*}

In turn, making progress on representative selection mechanisms has also direct benefits for their applications in computer systems. For example, some blockchains select validators via a \textit{nominated proof-of-stake} protocol, and the representativeness of the selection is essential for the security of the system \citep{CeSt21a}.  
Further afield, blockchain-enabled Decentralized Autonomous Organizations (DAOs) are at the forefront of testing innovative governance systems based on interactive procedures \cite{kashyap2020competent,anshelevich2021representative,zhang2017brief}. Good (e-)governance remains a vast open question \cite{hassan2021decentralized}.

\section{Mathematical Framework}\label{sec:mathframe}
In this section, we outline a mathematical framework to model  selection mechanisms. 
First, we define some preliminary notation. 
A matrix is stochastic if each row sums up to $1$. For a natural number $n\in \mathbb{N}$, let $[n]$ denote the set $\{1,2,\dots,n\}$ and let $e_n\in \mathbb{N}^{1\times n}$ denote the row vector containing all ones. 
 For a vector $a\in \mathbb{R}^{1\times n}$, let $\lVert a \rVert_1$ denote the $\ell_1-$norm of $a$, i.e., $\lVert a \rVert_1=\sum_{i=1}^{n} |a_i|$.

\subsection{Modeling Representation}
We present a mathematical framework for the following task: 
A group  $\N=[n]$ of $n$ agents wants to select a subset of $\N$ to act as a representative body through a selection mechanism~$M$. 
We additionally assume that the agents selected to be part of the representative body can have different voting weights, i.e., in a decision made by the representative body, some agents' votes have more weight than others.
Formally, given $\N$, we want to select a \emph{weight vector} $\mathbf{w}\in \mathbb{R}_{\ge 0}^{n}$. For each $i\in \N$, if $\mathbf{w}_i>0$, then~$i$ is selected as part of the representative body and has voting weight $\mathbf{w}_i$.
The size of the induced representative body is given by~$|\{ i \in \N \mid \mathbf{w}_i > 0 \}|$.

\paragraph{Representation Matrix.} The relation of agents is captured by a \emph{representation matrix} $\Gamma\in  \mathbb{R}^{n\times n}$, where the entry $\Gamma_{ij}$ describes how well agent $j$ can represent agent $i$. $\Gamma$ is a stochastic matrix that allows fractional entries to account for the fact that agent $i$ may be best represented by a mixture of other agents. How well $i$ feels represented by $j$ may be based on complex interactions of multiple aspects such that the issues $i$ cares about, the relative preferences of $j$ and $i$ on these and other issues, intrinsic characteristics of $i$ and $j$, and the underlying social network capturing who knows who.\footnote{In line with this reasoning, political theorists have argued that quality of representation is multidimensional and depends on different factors such as similarities between the representative and the constituents (\textit{descriptive representation} \cite{mansbridge1999should}), alignment of interests and values (\textit{gyroscopic representation} \cite{mansbridge2009selection}), or advancement of constituents' interests by the representative (\textit{substantive representation} \cite{pitkin1967concept}).}
Hence, the matrix $\Gamma$ captures the complex nature of  potential representation between agents in a simple yet rich form.

\begin{example}\label{ex:running}
    Consider the following example: $\N = \{A, B, C, D, E\}$ such that $A$ and $B$ belong to some party, and $C, D$ and $E$ to another party.
    Imagine that $A$ and $E$ are extreme candidates seeking power in their respective parties. Moreover, $B$ and $D$ are completely partisan and would never want to be represented by someone outside their parties. In contrast, $C$ is moderate in their beliefs and could be represented by other candidates with non-extreme views. This situation could be represented by the representation matrix $\Gamma$ given in Figure \ref{fig:gamma}. 

\end{example}

\begin{figure}
\[
\kbordermatrix{ & A & B & C & D& E\\
A& 1 & 0 & 0 & 0 & 0 \\
    B& 2/3 & 1/3 & 0 & 0 & 0  \\
    C& 0 & 1/3 & 2/3 & 0 & 0  \\
    D& 0 & 0 & 2/5 & 1/5 & 2/5 \\
    E& 0 & 0 & 0 & 0 & 1
	}
 \]
 \vspace{-1em}
    \caption{Representation matrix $\Gamma$ for the instance described in Example \ref{ex:running}. Rows and columns are indexed with agents.}
    \label{fig:gamma}\vspace*{-0.5cm}
\end{figure}

\paragraph{Interpreting the Representation Matrix.} The representation matrix can be interpreted as giving rise to voting behavior. Specifically, assuming that each agent can split their vote in an arbitrary way, the matrix entries can be thought of as the ideal split of an agent's vote into fractional votes. In this paper, we will focus on uninominal ballots, i.e., each agent can vote for exactly one other agent to be part of the representative body. 
Accordingly, we interpret the entry $\Gamma_{ij}$ as the probability that agent $i$ selects (i.e., votes for) agent $j$.\footnote{The translation of $\Gamma$ to votes can be extended to other ballot formats such as approval ballots or ranked ballots.} 

\paragraph{Expected Vote Share.} 
Using the probabilistic interpretation of~$\Gamma$ for uninominal ballots allows us to reason about the expected share of votes an agent receives. 
For this, let $V_j$ be the random variable representing the vote share agent $j$ receives under the representation matrix $\Gamma.$ Let $V$ be the vector of the $n$ random variables $V_1,\dots,V_n$. Then, the expected vote share $\mathbb{E}[V_j]$ of agent $j$ is the sum of the $j$-th column of $\Gamma$, i.e., $\mathbb{E}[V_j]=\sum_{i=1}^n \Gamma_{ij}$ and $\mathbb{E}[V] = \Gamma^Te_n$. 
In an idealistic setting, we would select all agents as members of the representative body and give each agent $j$ a voting weight of $\mathbb{E}[V_j]$, i.e., $\mathbf{w}_j=\mathbb{E}[V_j]$ for all $j\in \N$. 
Accordingly, to evaluate the quality of different selection mechanisms, we will compare the (ideal) expected vote share $\mathbb{E}[V]$ to the voting weights of agents returned by the mechanism, assuming that agents vote as described by $\Gamma$. For the representation matrix given in Figure \ref{fig:gamma}, the vector of expected vote shares of the agents is $\mathbb{E}[V]= \left(\frac{5}{3},\frac{2}{3},\frac{16}{15}, \frac{1}{5}, \frac{7}{5}\right)$.

\paragraph{Properties of the Representation Matrix.}
We envision that the algebraic properties of a given representation matrix $\Gamma$ could model salient societal characteristics of $\N$ relevant to an axiomatic analysis: polarized groups would be characterized by a block matrix $\Gamma$, the relative magnitude of $\Gamma$'s trace would quantify the amount of power-seeking agents in the group, the rank of $\Gamma$ would model how correlated agents are to each other, etc. In line with axiomatic results from social choice theory \citep{ASS02a,Gaer01a}, we expect to find impossibility theorems that can be circumvented by restricting the structure of~$\Gamma$. 

\subsection{Selection Mechansisms}
We want to analyze selection mechanisms $M$ that, given the uninomial ballots from the agents, return a weight vector.
As an additional part of the input, our mechanisms may take a pre-specified subset $\C \subseteq \N$ of $m$ agents acting as \textit{candidates} and an integer $k$ describing the number of agents that can be selected to be part of the representative body.
For a mechanism $M$ and body size $k$, we define a function $f^{M_k}$ that given $\Gamma$ and $\C$ returns the candidates' expected voting weights under $\Gamma$ and $\C$.\footnote{Some mechanisms need neither a candidate set $\C$ nor the size of the representative body $k$ as input (in this case we drop $k$ from $f^{M_k}$).}  
Formally, $f^{M_k}$ is a function
\[
    f^{M_k}: \mathbb{R}^{n\times n} \times (2^{\N} \setminus \emptyset) \to \mathbb{R}^{n\times 1}
\]
such that $\{i \in \N \mid f^{M_k}(\Gamma, \C)_i>0\}$ is a subset of $C$ of size at most $k$. 
Here, $f^{M_k}\left(\Gamma, \C \right)_i$ is the expected voting weight ($\mathbb{E}[\mathbf{w}_i]$) of candidate $i\in \C$ in a body of size $k$ selected by mechanism $M$ assuming that $i\in \N$ votes for $j\in \C$ with probability depending on $\Gamma_{ij}$. 

We now describe how the expected voting weights for different selection mechanisms can be computed, using the representation matrix given in \Cref{fig:gamma} as a running example. 

\subsubsection{Direct Democracy (D)}
    In direct democracy assemblies,
    all agents are elected in the represented body. Thus, $\C = \N$ and $f^{D}\left(\Gamma, \N \right) = e_n$ for all representation matrices $\Gamma$. 
    

\subsubsection{First-Past-The-Post (F)} 
First-past-the-post voting is  widely used around the world but also widely criticized  for, among other things, leaving voters feeling underrepresented \cite{bogdanor1997first}. In first-past-the-post, a voting weight of $1$ is given to the candidate receiving the highest number of votes and $0$ to all other candidates.
Notably, in first-past-the-post elections the electorate is typically partitioned into different voting districts, each selecting its own representative. 
We focus on the single-district case; however, our model can be extended to parallel independent districts.

Continuing \Cref{ex:running}, let $\C = \{A, B, C, E\}$.  Note that the function $f^{F_1}$ alters the representation matrix to account for the set of candidates:
agents can only vote for candidates, and we assume that candidates always select themselves.  In the running example, the representation matrix projected on the set of candidates becomes 
$$
\kbordermatrix{ & A & B & C & D& E\\
A& 1 & 0 & 0 & 0 & 0 \\
    B& 0 & 1 & 0 & 0 & 0  \\
    C& 0 & 0 & 1 & 0 & 0  \\
    D& 0 & 0 & 1/2 & 0 & 1/2 \\
    E& 0 & 0 & 0 & 0 & 1
	}.
 $$
With probability $0.5$,
$C$ receives $2$ votes and $A, B$ and $E$ receive $1$ vote (resulting in $C$ having a voting weight of $1$ in the representative body and all other agents having a voting weight of zero), and with probability $0.5$, $E$ receives $2$ votes and $A, B $ and $C$ receive $1$ vote. 
Consequently, we get $f^{F_1}(\Gamma, \C ) = \left(0, 0, 1/2, 0, 1/2\right)^T$.

\subsubsection{Proxy Voting (P)}
In proxy voting, all agents are presented with a pre-defined pool of candidates, and each agent can delegate their voting power to one of the candidates. All candidates are \emph{de facto} part of the representative body, and candidates have a voting power proportional to the number of votes delegated to them.\footnote{Proxy voting is closely related to the widespread practice of \textit{party-list elections} \citep{Puke14a}, where agents vote for parties and the seats in the representative body are distributed so that the number of seats of a party is proportional to the number of received votes. We focus on proxy voting as it allows for a cleaner mathematical formulation.}
Proxy voting has been studied both within computer science \cite{cohensius2017proxy,anshelevich2021representative} and political sciences \cite{miller1969program}, with some works extending it to more flexible issue-based delegations \cite{abramowitz2019proxy}.  

The expected voting weight under proxy voting is the sum of expected delegations for the proxies  (as in First-Past-The-Post, we adapt the representation matrix to account for the candidate set). Assuming again $\C = \{A, B, C, E\}$ in \Cref{ex:running}, since $D'$s vote goes to $C$ with probability $0.5$ and to $E$ with probability $0.5$, we get $f^{P}(\Gamma, \C) = \left(1, 1, 3/2, 0, 3/2\right)^T$. 

\subsubsection{Liquid Democracy (L)}
In liquid democracy, each agent can choose to be part of the representative body or delegate their vote to another agent. Delegations are transitive, i.e., if $A$ delegates to $B$ and $B$ delegates to $C$, and $C$ decides to be in the representative body, then $C$ votes on behalf of themself, as well as $A$ and $B$. The representative body consists of all agents who self-select, with their voting power being set to the number of votes (transitively) delegated to them plus one. 
Liquid democracy has received considerable attention in the computer science community by studying it from a procedural and epistemic perspective \cite{kahng2021liquid,escoffier2019convergence,bloembergen2019rational,zhang2021power,halpern2021defense,brill2022liquid,golz2021fluid,revelliquid}, developing dedicated supporting software \cite{behrens2014principles,paulin2020overview}, and examining possible extensions \cite{ford2002delegative,colley2022unravelling,brill2022liquid}. Liquid Democracy has also been scrutinized from a political science perspective \citep{miller1969program,blum2016liquid,valsangiacomo2022clarifying}.

To find the expected voting weights of the agents under transitive delegations, we leverage the representation matrix to exhaust all possible configurations of transitive delegations and compute their probability. For instance, one possible configuration in \Cref{ex:running} is that every agent votes for themselves (which happens with probability $\frac{2}{45}$), resulting in all agents being part of the representative body and having voting weight $1$.
\footnote{Note that if there is a delegation cycle, the votes of agents in the cycle are lost. Accordingly, their voting weight is set to zero and the agents are effectively ignored.} Overall, the expected voting weights are as follows:
$f^{L}(\Gamma, \N ) = \left(\frac{89}{45}, \frac{22}{45}, \frac{14}{15}, \frac{1}{5}, \frac{7}{5}\right)^T$.
\subsubsection{Sortition (S)}
Sortition is a selection method which draws at random a subset of the population to act as the representative body \cite{mueller1972representative,dowlen2017political,landemore2020open,flanigan2020neutralizing}. The method allows equal access to decision-making and does not require a voting phase. Agents who do not participate in the representative body despite being selected pose problems with the fairness guarantees offered by sortition. Computer scientists are investigating algorithmic ways to deal with this issue \cite{flanigan2020neutralizing,DBLP:journals/nature/FlaniganG0HP21}. 
The representative body to be found has a fixed size $k < n$ and is found uniformly at random from $\N$. All members of the representative body have an equal voting weight. Thus, the expected voting weight of each agent is $\frac{k}{n}$, i.e., $f^{S_k}\left(\Gamma, \N \right)=\frac{k}{n}e_n$.

\medskip 
Note that selection mechanisms differ with respect to different dimensions, in  particular, \begin{enumerate*}[label=(\roman*)]
\item whether candidates are pre-selected ($m<n$) or anyone can be part of the representative body ($m=n$),
\item whether the output representative body has a predefined size or not, and  
\item whether each agent has a direct link to some member of the representative body they support ($\lVert f^{M_k}\left(\Gamma, \C \right)\rVert_1 = n$), or some agents are virtually represented (by someone they did not necessarily vote for) ($\lVert f^{M_k}\left(\Gamma, \C \right)\rVert_1 < n$).
\end{enumerate*}
We call these dimensions \textit{open-closed}, \textit{flexible-rigid}, and \textit{direct-virtual}, respectively. The above-described selection mechanisms are all located on different positions of the induced 3-dimensional space because some allow more flexibility, or represent more agents by design. We want to understand the impact of these design choices on desirable axioms.  In turn, the arising 3-dimensional space helps with comparing different mechanisms. We envision that mechanisms from a certain region of this 3-dimensional space perform particularly well (or not) with respect to some of our axioms.

\section{Axioms}
We focus on five axioms, 
each capturing different aspects of representation: proportionality, diversity, monotonicity,  faithfulness, and effectiveness. 
We lean on both the field of (computational) social choice and political science for these axioms.
Our described axioms are a first step toward clarifying 
what various selection mechanisms entail; this is not to pretend that these desiderata are the only ones that matter or that they are the ``most desirable.'' For instance,  one could want to study the selection mechanisms with respect to the quality of decisions made by the selected body or the accountability of the selected body. This, however, is outside the scope of our paper.
\paragraph{$\varepsilon-$proportionality} Proportionality captures how ``accurately'' the expected voting weights of agents in the representative body reflect their expected vote share. Proportionality is particularly desirable to achieve descriptive representation \cite{black1958theory,brams,mansbridge1999should,valsangiacomo2021political,zick2013random}, and relates to previous investigations in political science on proportionality metrics for different selection formulas \cite{rae1967political}.\footnote{Note that proportionality can be defined as the descriptive representation of votes, attributes, or preferences \cite{mansbridge1999should}. Here, we only model the proportionality of expressed votes rather than any other characteristic of the electorate.} We give a notion of $\varepsilon$-proportionality which insists for each candidate that their normalized expected vote share and  normalized expected voting weight differ by at most $\varepsilon.$ To define this, let $$\mathit{diff}(\Gamma,\C,M_k)= \max_{j\in [n]} \left|\frac{\mathbb{E}[V_j]}{\lVert\mathbb{E}[V]\rVert_1}-\frac{f^{M_k}\left(\Gamma, \C \right)_j}{{\lVert f^{M_k}\left(\Gamma, \C \right)\rVert_1}}\right|.$$
Then, $\varepsilon$-proportionality requires that 
$\mathit{diff}(\Gamma,\C,M_k)\leq \varepsilon$.
For selection mechanisms that depend on a closed set of candidates, define $\overline{\varepsilon^{M_k}} = \max_{\C \subset \N} \mathit{diff}(\Gamma,\C,M_k)$ and $\underline{\varepsilon^{M_k}} = \min_{\C \subset \N} \mathit{diff}(\Gamma,\C,M_k)$ as the maximum, respectively minimum, largest deviation of a candidate's expected voting weight from its expected vote share over all possible candidate sets.

To give an example for $\varepsilon-$proportionality, we again make use of the setting described in \Cref{ex:running}. 
In Table~\ref{tab:prop}, we give the values of $\varepsilon$ for which each of the selection mechanisms are $\varepsilon$-proportional on \Cref{ex:running}.  We see that liquid democracy has the smallest value of $\varepsilon$ in this case; whereas first-past-the-post does not manage to distribute the voting weight to the selected body as efficiently. An interesting takeaway from studying the selection mechanisms via this axiom is that we see that the proportionality of sortition is independent of the size of the body, including the case where the sortition is the size of the population ($k=n$) and direct democracy is recovered.

\paragraph{Diversity} Mostly relevant in the deliberation stage \cite{chamberlin1983representative,landemore2020open,duarte2015political}, we interpret diversity  as requiring that all opinions should be present in the representative body. 
We formalize it as ``if the expected vote share of a candidate is positive, then so should be their expected voting weight'', i.e., 
$\mathbb{E}[V_j]>0$ implies $f^{M_k}\left(\Gamma, \C \right)_j>0$ for all $j \in \C$.

\paragraph{Monotonicity} This benchmark is standard in social choice theory \citep{moulin1991axioms}.
Let $\Gamma$ and $\Gamma'$ be some representation matrices. If in the representation matrix $\Gamma'$ the expected vote share of a candidate $j$ is larger than in $\Gamma$ and the expected vote share does not increase for any other candidates, then $j$'s expected voting weight increases.  
That is, if $\Gamma$ and $\Gamma'$ be such that $\mathbb{E}[V'_j] > \mathbb{E}[V_j] \text{ and } \mathbb{E}[V'_i] \le \mathbb{E}[V_i] \text{ for all } i\neq j$, then the inequality $f^{M_k}(\Gamma', \C)_j \geq f^{M_k}(\Gamma, \C)_j$ should hold.

\paragraph{Faithfulness}
This axiom ensures that candidates are not hurt by having a higher expected vote share. The axiom requires that if a candidate has a higher vote share than some other candidate, then they also have a higher expected voting weight as computed by the mechanism $M$, i.e., $\mathbb{E}[V_i] \geq\mathbb{E}[V_j]$ implies $f^{M_k}(\Gamma, \C)_i\geq f^{M_k}(\Gamma, \C)_j$ for all $i,j\in \C$. 

\paragraph{$\gamma-$effectiveness} Finally, \textit{effectiveness} models potential deadlocks when no majoritarian coalition may come to an agreement.
This benchmark measures the size of the smallest coalition needed to have majority support for some proposal. For a given mechanism $M$ and candidate set $\C,$ it is defined as the expected smallest number $\gamma^{M_k}_{\C}$  such that some coalition of $\gamma^{M_k}_{\C}$ representatives gather strictly more than half of the voting weight. For mechanisms that rely on a specified set of candidates, it would again be interesting to look at the worst and best-case scenarios for $\gamma^{M_k}_{\C}$ over all possible candidates set of fixed size. 

\begin{table}[]
\begin{tabular}{p{0.5cm}p{1cm}ccp{1cm}p{1cm}}
\toprule
& \phantom{x}$f^{D}$ & $f^{F_1} $ & $f^{P}$ & \phantom{x}$f^{L}$ & \phantom{x}$f^{S_k}$  \\
\midrule
$\varepsilon$   & 0.16 & [0.33, 0.86] & [0.13, 0.33]& 0.06 & 0.16 \\
\bottomrule 
\end{tabular}
\vspace{0.5em}
\caption{The minimum values of $\mathbf{\varepsilon}$ for $\mathbf{\varepsilon-}$proportionality in \Cref{ex:running}  for each of the different selection mechanisms: direct democracy (D), first-past-the-post (F), proxy voting (P), liquid democracy (L), and sortition (S). The intervals 
denote the best and worst-case $\mathbf{\varepsilon}$ over all candidates sets. For sortition, the presented $\epsilon$ value holds for all sizes of the body~$\mathbf{k\in [n]}$.
\vspace*{-0.5cm}
}\label{tab:prop}
\end{table}

\section{Discussion}
We have argued that there is a need for a more systematic comparison of different selection mechanisms within a unified framework to understand the trade-offs inherent to the selection mechanisms currently on the table to open democratic representation \cite{landemore2020open}.  
Taking a first step in this direction, we have presented a simple model that allows the formulation of many different selection mechanisms together with axioms derived from political science and social choice theory that can be used to compare and assess these mechanisms.

We do not see our model and axioms as final or exhaustive, and we believe that asking the right questions is already the first research challenge. 
Nevertheless, there are interesting open questions arising from our study:
Which of these mechanisms always satisfy diversity, monotonicity, and faithfulness? Can we obtain meaningful bounds on the $\epsilon$-proportionality or $\gamma$-effectiveness of the different mechanisms? While it seems unlikely that general bounds can be obtained, we hope that identifying characteristics of the representation matrix could correspond to a guarantee of certain proportionality and effectiveness values.
Moreover, it would be interesting to obtain comparative statements in the sense that one mechanism is always guaranteed to be better than another (at least if the ``right'' candidate set is chosen). 
Lastly, one may also wonder about the influence of the number of candidates and the selected set of candidates on the axiomatic performance of our mechanisms.
More generally, it would be interesting to identify general characteristics of the selection mechanisms that benefit axiomatic guarantees,
potentially extending upon our discussed \textit{open-closed}, \textit{flexible-rigid}, and \textit{direct-virtual} dimensions.

\newpage
\bibliography{biblio}


\end{document}